%Paper: hep-th/9504120
%From: ANSAR@nbivax.nbi.dk
%Date: Mon, 24 Apr 1995 16:31:12 +0200

\input harvmac

%For more complicated situations, substitute for {\it either\/} argument:
\Title{\vbox{\baselineskip12pt\hbox{NORDITA 95/22}}}
{\vbox{\centerline{Some Comments on $N=2$}
	\vskip2pt\centerline{Supersymmetric Yang-Mills}}}
%   \footnote{}{*optional footnote on title}

\centerline{Ansar Fayyazuddin \footnote{$^\dagger$}
{ansar@nbivax.nbi.dk}}
\bigskip\centerline{NORDITA}
\centerline{Blegdamsvej 17}\centerline{DK-2100 Copenhagen, Denmark}

%if too many authors for abstract on same page, say   \vfill\eject\pageno0

\vskip .3in
We comment on some aspects of the semiclassical BPS saturated states
close to the curve ${\rm Im} a_{D}/a = 0$.
\Date{3/95} %replace this line by \draft  for preliminary versions
	     %or specify \draftmode at some point

%if you want double-space, use e.g. \baselineskip=20pt plus 2pt minus 2pt

Recently Seiberg and Witten have given an exact description of the low-energy
$N=2$ Yang-Mills theory \ref\rSW{N. Seiberg and E. Witten, Nuc. Phys. B426
(1994) 19, hep-th/9407087\semi Nucl. Phys. B431 (1994) 484, hep-th/9408099}.
 They have furthermore determined the exact mass
spectrum of the BPS saturated states in the theory.  More precisely
if one knows the
quantum numbers (in this case the magnetic and
electric charges) of a particle which exists then one can calculate its
exact mass.
The spectrum of states in the semi-classical limit is known to consist
of the $N=2$ vector multiplets with the photon and electrically charged
$W$s as their spin $1$ components.
There are also solitons carrying magnetic charge $1$ and arbitrary
integral electric charge.  The magnetically charged solitons form
spin $\leq 1/2$ hypermultiplets \ref\rOsb{H. Osborne, Phys. Lett. B83 (1979)
321.}.
All the semiclassical states
saturate the so-called BPS bound.
The quantum theory has a space of vacua which is parameterized
by $u=<tr\phi^{2}>$, where $\phi$ is the scalar component of the
$N=2$ vector multiplet.  The low-energy theory
is $N=2$ supersymmetric QED almost everywhere: at two
points (which can be taken to be $u=-1,1$) extra particles become
massless, these particles have spin $\leq 1/2$
and so the $SU(2)$ gauge symmetry is never restored.
As argued in \rSW\ the semi-classical BPS saturated
states continue to exist as long as one does not cross a curve on
which $a_{D}/a$ is real.  When such a curve is crossed some of the
states which exist semi-classically become degenerate with some multi-particle
stable BPS states.  In this note we demonstrate : 1)
that there is a curve passing through $u=-1,1$ on which $a_{D}/a$ is real
and that this curve
is diffeomorphic to a circle; 2) that all the states except for the
monopoles/dyons becoming massless at $u= -1, 1$ become degenerate with
multiparticle
states consisting of only these two particles on the ${\rm Im}a_{D}/a=0$
curve.  Further, assuming that the only semi-classically massive BPS
saturated particles which continue to exist are these two distinguished
states, then the monodromies naturally work out.

We first demonstrate that there is a curve diffeomorphic to the
circle on which $a_D/a$ is real.  Notice that this condition is
an SL(2,Z) invariant statement, so it is natural to find an SL(2,Z) invariant
{\it function} on the $u$ plane which takes a particular value
when and only when this condition is satisfied.  The K{\" a}hler potential
$$K = {\rm Im}{\bar a}{a_{D}}$$ is such a function.
$K=0$ if and only if ${\rm Im} a_{D}/a =0$.
The manifold we are interested in is then $N = K^{-1}(0)$.
If we can show that $0$ is not a critical value of $K$ then
we will have shown that $N$ is a smooth one dimensional manifold
\ref\rmilnor{J. W. Milnor, ``Topology from the differentiable viewpoint'',
The University Press of Virginia, Charlottesville (1965), p. 11.}.  Once we
have established this fact we need only check that the $K=0$ curves
passing through $u=-1,1$ are not disjoint and do not continue to infinity.

We begin by establishing that $0$ is not a critical value of $K$.
Let us fix a point $u \neq -1,1$.  Then $dK = 0$ and $K=0$
are both satisfied only if ${\rm Im} \partial_{u} a_{D}/\partial_{u}a =0$.
However, we know that $K$ is a good K{\" a}hler potential everywhere
except at $u={-1,1}$, therefore this condition is in conflict with
a positive metric and cannot be satisfied away
from the points $u=-1,1$.  Now let us take $u=1$, then using the explicit
expressions calculated in \rSW\ we know that $dK = 1/\pi (du + d{\bar u})
\neq 0$ at $u=1$.  Similarly $dK\neq 0$ at $u=-1$ either, this follows
from the $Z_{2}$ symmetry of the theory.  Explicitly,
$a(u)= -i4/\pi +i/2\pi (u+1)\ln (u+1)$ and
$a_{D}+a = 1/2(u+1) + O(u+1)^2$
close to $u=-1$.  Therefore
we see that $K^{-1}(0)$ is a smooth one dimensional manifold \rmilnor .

We can now eliminate the possibility that either one
of the $K=0$ curves passing through $u=-1,1$ continue to infinity.
This follows from the expressions evaluated in \rSW\ for $\mid u\mid >>1$:
$K\approx 2/\pi\mid u\mid\ln\mid u\mid > 0$.
There are now only two possible behaviors\foot{We are making use of the fact
that at $u=-1,1$ the
real axis is not tangent to the curve $K=0$.  This has been shown
for $u=1$ in \rSW\ and for $u=-1$ follows from the $Z_{2}$ symmetry
of the theory and from the explicit expressions given for $a_{D},a$
in the previous paragraph.}
for the $K=0$ curves
passing through $u=-1,1$:
1)that they form two disjoint closed curves or 2)that there
is a curve diffeomorphic to the circle passing through $u=-1,1$.
The first scenario is easily
seen not to be realized from the expressions for $a_{D}, a$
in terms of integrals given
in \rSW\ , one sees
immediately that for $u$ real and $-1 < u < 1$ , ${\rm Re} a_{D}=0,
{\rm Im} a_{D}\neq 0$
and ${\rm Re} a\neq 0, {\rm Im} a\neq 0$.
Similarly for $u$ real and $u <-1$ we
have ${\rm Im} a_{D} \neq 0, {\rm Re} a_{D}\neq 0$ and
${\rm Im} a\neq 0, {\rm Re} a=0$.  The situation for $u$ real and $u>1$
has already been discussed in \rSW\ where it was shown that $K\neq 0$ in
that region.  So the
only points on the real axis where $K=0$ are $u=-1, 1$.
This establishes that there is a curve diffeomorphic to $S^{1}$ passing
through $u=-1,1$.  An analytic argument for the existence of this curve
has also been given in\foot{I would like to thank J. de Boer for
informing of this reference.}\ref\rargyres{ P. Argyres, talk given
in Los Angeles, 1995 (J. de Boer private communication)}.

Since the manifold $K^{-1}(0)$ consists in general of (possibly) many
disjoint one dimensional manifolds diffeomorphic to $S^1$ we shall
denote by $\tilde N$ the particular one passing through the points $u=-1,1$.
We would now like to discuss the physical spectrum
as we cross the curve $\tilde N$.  We begin by establishing
some general features.  Let us first recall that according to \rSW\
there are only two points on the $u$-plane at which two semiclassically
massive states become massless.  This implies the following: $a_{D}/a$
is an integer only at $u=-1, 1$, since if at some point
$a_{D}/a =n$ then the state $(1,-n)$ becomes massless at this point.
Therefore, on the curve $\tilde N$ $a_{D}/a$ is non-integer except at
the points $u=-1,1$ where it is integral.  The values of $a_{D}/a$
at these two points determine in which range the values of $a_{D}/a$ will
take.
	Let us consider a monodromy around a point $u=u_{0}$
with a base point in the
semi-classical region ($\mid u\mid >>1$) and located in the
$u$ upper-half plane.  If the particle becoming massless at $u=u_{0}$ has
charge vector $(1,n)$ then the monodromy matrix for an anti-clockwise
path around $u_{0}$ is given by:
$$M_{u_{0}}=(\matrix{1+2n & 2n^{2}\cr -2 & 1-2n \cr})$$.
If we restrict ourselves
to a region including the base point and $u_{0}$ but no other singular
points.  Then in this region $a_{D}+na$ is a good coordinate
and has a Taylor expansion close to $u_{0}$.  Although $a$ does not
have a Taylor expansion, it can be expanded in fractional powers
of $(u-u_{0})$ and $(u-u_{0})^{k}\ln (u-u_{0})$.  Specifically,
$$a_{D}+na \approx c(u-u_{0}) + o(u-u_{0})^{2} $$
$$a \approx b + i{{c}\over{\pi}} (u-u_{0})\ln (u-u_{0})$$
where $b,c$ are constants.  If we also assume that $\tilde N$ looks
like $u= u_{0}+it$ for $t$ real close to $u=u_{0}$, then ${\rm Re} c/b=0$.
We come to the main point of this detour which is to fix the relative
sign of $a_{D},a$ which individually do not have well defined signs.
The relative sign appears in the masses of the BPS saturated states
and is therefore physical.  The curve $\tilde N$ is a smooth manifold
and we know that there is a path to $u_{0}$ from the semi-classical
region along which ${\rm Im} a_{D}/a >0$, in particular approaching
$u=-1(1)$ from $-\infty (+\infty)$ along the real axis never crosses
a point where ${\rm Im} a_{D}/a =0$.  $\tilde N$ defines a region
which is essentially the strong coupling region.  Immediately outside
(resp. inside) of this region ${\rm Im}a_{D}/a >0$
(resp. ${\rm Im} a_{D}/a <0$).  This fixes the sign of ${\rm Im} c/b$ as
follows, if $u_{0}=-1$ then ${\rm Im} c/b <0$ and if $u_{0}=1$ then
${\rm Im} c/b >0$.
	Close to $u=u_{0}$, $$a_{D}/a \approx -n + {{c}\over{b}} (u-u_{0}),$$
so $a_{D}/a$
increases if one follows $\tilde N$ in a clockwise direction and decreases
in the anti-clockwise direction.  In fact one can say more: $a_{D}/a$
is monotonic on $\tilde N$.  This can be established by noting that
the derivative of $a_{D}/a$ is always non-zero on $\tilde N$ if the
metric is positive everywhere on $\tilde N$ (excluding the two singularities)
and assuming that $a$ never becomes infinite.

Before proceeding to the
specifics, we note a potentially confusing point.  At $u=u_{0}$
the lowest mass particle is the one becoming massless at that point,
i.e. a particle with charge $(1,n)$.  The particles with the next highest
mass are the ones with charges $(1,n-1), (1,n+1), (0,1)$.  These three
particles are degenerate in mass, the first two are dyons and the last
the $W$.  As we move away from $u=u_{0}$ the degeneracy in mass is lifted,
however the identity of the lightest particles is different depending
on whether one moves clockwise or anti-clockwise from $u_{0}$ on
$\tilde N$.  While
$(1,n)$ always remains one of the two lightest particles, the other
light particle is $(1,1+n)$ (resp. $(1, n-1)$) if one moves anti-clockwise
(resp. clockwise) away from $u_{0}$ on $\tilde N$.  If one follows
the curve $\tilde N$ towards the singularity lying outside of the
region enclosed by the path defining the monodromy one will see
different particles becoming massless at this point.  This is consistent
since the identity in terms of charge vectors of the particle becoming
massless at the other singularity is not invariant under the monodromy
$M_{u_{0}}$.  However, presently we will propose that only the
two lightest dyons be present inside the strong coupling regime and
we will label them differently depending on whether we are considering
a clockwise or anti-clockwise path around $u_{0}$.
That this proposal is plausible can be seen as follows.  If one asks
using the monodromy $M_{u_{0}}$ how many particles come back with a
mass equal to that of a semi-classical state the answer is precisely two.
These two states are $(1,n)$ and $(1,n+1)$ for the anti-clockwise
monodromies and $(1,n)$ and $(1,n-1)$ for the clockwise monodromies.

Consider a closed path based at a point located at $\mid u\mid >>1$
(i.e. in the semi-classical region) with ${\rm Im} u>0$
 which wraps once around the
point $u=1$.  The path is homotopic to a path which crosses
$\tilde N$ once in the upper and once in the lower-half planes.
As the path crosses $\tilde N$ in the upper half plane
$-1<a_{D}/a<0$ as shown above.  The lowest mass
particles at the point where the path crosses $\tilde N$ are
$(1,0), (1,1)$, with masses $\sqrt{2}a_{D}$ and
$\sqrt{2}(\mid a\mid - \mid a_{D}\mid )$.
All the other semi-classical states have masses
and charges degenerate with a multi-particle state consisting
solely of these two particles.  In particular, the $W$s with
charge vectors $(0, \pm 1)$ have the same mass as the sum
of the masses of the particles $(\mp 1,0)$ and $(\pm 1, \pm 1)$.
Since the charges also add up the $W$s must be degenerate with a
two particle state consisting of these two particles.
Similarly, the state $(1, n)$ with $n \geq 2$
has mass equal to the sum of the masses of $n-1$ monopoles
with charge $(-1,0)$ and $n$
dyons with charge $(1,1)$.  Again since the charges add up the
state $(1,n)$ is degenerate with a $2n-1$ particle state consisting
solely of the two particles becoming massless at $u=-1,1$.
The remaining states with charge vector $(1,-n)$ with $n\geq 1$
are degenerate with a $2n+1$ multiparticle state consisting of
$n$ dyons of charge $(-1,-1)$ and $n+1$ monopoles of charge $(1,0)$.
Since the states $(1,1),(1,0)$ do not become unstable as one enters
the strong coupling regime there must exist one-particle states with
these quantum numbers.  The other semi-classical dyons are
related to these two states by an $M_{\infty}$ transformation.
Since an $M_{\infty}$ monodromy based at a point in the strong coupling
regime always has to cross $\tilde N$, it is not necessary for the
other dyons to exist as one-particle states inside the strong coupling regime.
Also, the $W$ need not exist either.  If we assume this rather ``maximal''
scenario (i.e. the only semi-classical states which survive
in the strong coupling regime are the
particles becoming massless at $u=-1,1$), then as we cross the
$\tilde N$ in the lower half plane all the semi-classical states
return as multi-particle states except for $(1,1)$ which transforms to
$(-1,1)$ and $(1,0)$ which returns untransformed.

Similarly, a clockwise path will be homotopic to a path which
crosses the $\tilde N$ in the lower half plane first.  At the intersection all
the particles except $(1,-1)$ and $(1,0)$ will become degenerate with
a multiparticle state as follows: $$\eqalign{&(0,1)\rightarrow (-1,1) + (1,0)
\cr &(1,n) \rightarrow n(-1,1) + (n+1)(1,0)   \,\,
({\rm with }\,\, n\geq 1) \cr
&(1,-n)\rightarrow n(1,-1) + (n-1)(-1,0) \,\, ({\rm with } \,\, n\geq 2)}$$
If we again assume that the only semi-classical one-particle states in the
strong coupling region are $(1,0)$ and $(1,-1)$ (recall that the
identity of the stable particles is not the same for the
anti-clockwise and clockwise mondromies), then all the states
except for these two come back as multiparticle semi-classical states.
The state $(1,0)$ is not transformed but $(-1,1)$ comes back as
a $(1,1)$ dyon.

The monodoromies around $u=-1$ can be studied in a similar way.
The only difference being that the identity of the stable particles
in terms of charge vectors
will be different from the $u=1$ monodromies.  In particular, the
semi-classical particles which will return as one particle states
are $(1,1)$ and $(1,2)$ ($(1,0)$) for the anti-clockwise (clockwise)
monodromies.

In a recent paper \ref\rlr{U. Lindstr{\" o}m and M. R{\^ o}cek, ``A note
on the Seiberg-Witten solution of N=2 Super Yang-Mills Theory'', USITP-95-4,
hep-th/9503012.} Lindstr{\"o}m and R{\^ o}cek have noted that the
$W$ bosons have a negative kinetic term if one continues the
semi-classically correct effective action to strong coupling.
It would be interesting to devise methods of probing the spectrum
in the stong coupling regime to establish whether the scenario
presented here is infact realized.

The existence of the curve ${\rm Im}a_{D}/a =0$ has been established
numerically in \ref\curve{J. de Boer, D. Jatkar, C. Skenderis and B. Peters,
unpublished\semi P. Argyres and A. Faraggi, unpublished (as quoted in \rlr )}

\bigbreak\bigskip\bigskip\centerline{{\bf Acknowledgements}}\nobreak
I have benefited immensely from
discussions with F. Bastianelli, P. Di Vecchia, and P. Ernstr{\" o}m who
sharpened many of the issues discussed in this paper and
offered many constructive suggestions.  I would also like to thank
J. de Boer for an illuminating correspondence.
\listrefs
\bye